# Magnetic Properties of Potential Li-ion Battery Materials

Md Rakibul Karim Akanda, Amaya Alexandria Holmes, Jinorri Wilson

Department of Engineering Technology, Savannah State University—Savannah, GA 31404, United States of America

**Abstract**

Lithium-ion batteries (LiBs) have transformed electrochemical energy storage technologies and made a substantial contribution to grid-scale energy storage and the e-mobility revolution. Notwithstanding their many benefits, safety issues—specifically, thermal runaway incidents—have drawn attention from all around the world. In addition to discussing safety concerns, cooling techniques, and the history of battery materials, this study offers a thorough analysis of the growth, difficulties, and developments in Li-ion battery technology. Quantum Espresso software has been used to compute the magnetic characteristics of several potential Li-ion battery materials, which can enhance Li-ion battery performance.

1. **Introduction**

Because of its quick reaction time, bi-directional control, small size, and environmental adaptability, lithium-ion batteries are essential to electrochemical energy storage systems. Nonetheless, safety issues pertaining to thermal runaway events have drawn attention from people all around the world. Energy storage power plants throughout the world have seen 32 explosions connected to thermal runways in the last ten years. Both active and passive cooling techniques are used to address temperature problems in lithium-ion batteries. Phase change materials (PCM) or heat pipes are used in passive cooling, whereas air or liquid is used in active cooling. Combining PCM cooling with active techniques like liquid or air cooling is a popular strategy. Every approach has benefits and drawbacks. Although less effective at large heat flow densities, air cooling is still reasonably priced. Better heat transfer is possible with liquid cooling, but there are hazards of leaks and complicated systems needed. Heat pipes are quite effective in transferring heat, but they are heavy and complicated. To enhance thermal management, composite cooling approaches are becoming more and more popular. To improve review quality and predict future advances, bibliometrics is used to examine the research landscape in this area and provide insights into trends, authors, institutions, and research directions.

The reversible intercalation of Li+ ions into electronically conducting materials is how Li-ion batteries, also referred to as lithium-ion batteries, store energy. Compared to conventional rechargeable batteries, Li-ion batteries offer higher specific energy, higher energy density, higher energy efficiency, longer cycle life, and longer calendar life. Among all innovations, the invention and commercialization of Li-ion batteries are believed to have had one of the most significant societal consequences in human history, according to the 2019 Nobel Prize in Chemistry.

It is also widely used in grid-scale energy storage, in addition to military and aerospace applications. If lithium-ion batteries are not properly designed and manufactured, inappropriate handling or charging of the batteries could endanger public safety because they contain volatile electrolytes that can catch fire or explode if damaged. Significant progress has been made in the research and manufacturing of safe lithium-ion batteries. Lithium-ion solid state batteries are being developed to eliminate the flammable electrolyte. Inadequate recycling of batteries can result in hazardous waste and fire hazards, especially if the batteries include dangerous metals. In addition, there are significant obstacles in the extraction of strategically significant minerals like cobalt and lithium, which require a lot of water in often desert areas. Batteries make use of these minerals.

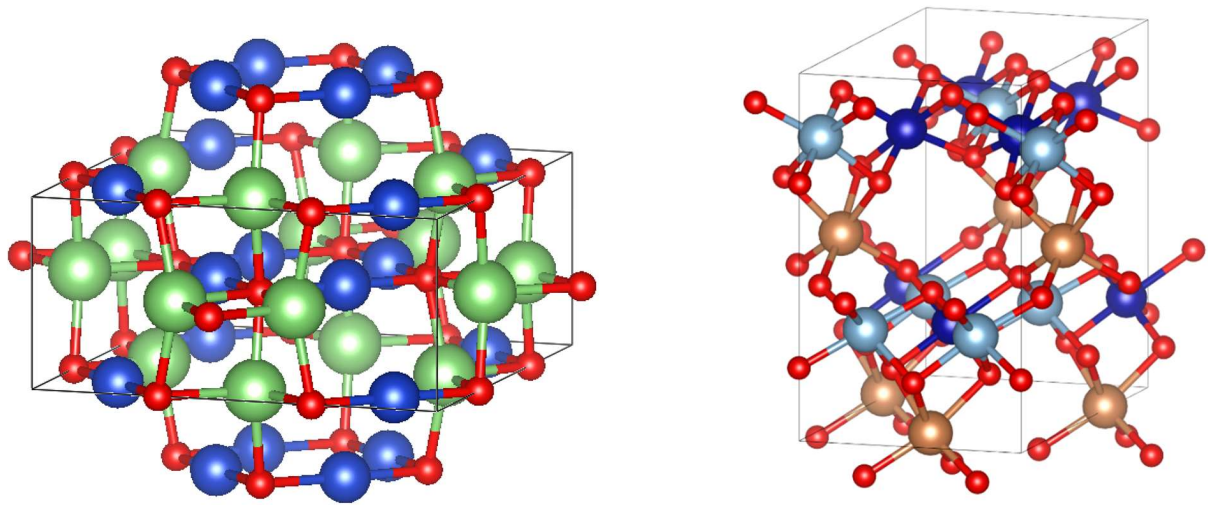

Figure 1. Crystaline Structure of LiCuO (mp-5127) and Al3Cr3-SbO8 (mp-756662)

NASA developed a CuF/Li battery in 1965, marking the beginning of research on rechargeable Li-ion batteries in the 1960s. The first Li-ion battery was created in 1974 by British chemist M. Stanley Whittingham, who used titanium disulfide (TiS2) as the cathode material. When Exxon tried to commercialize this battery in the late 1970s, they discovered that the synthesis was costly and complicated because of the susceptibility of TiS2 to moisture and the possibility of spontaneous fire. Over the past years several studies have been conducted to find the properties of materials using various software [1]. Consequently, Whittingham's lithium-titanium disulfide battery work was abandoned by Exxon [2]. TiS2 was substituted with lithium cobalt oxide (LiCoO2) in 1980 by Ned A. Godshall et al., Koichi Mizushima, and John B. Goodenough. LiCoO2 has a similar layered structure but is more stable in air and has a higher voltage. Although this substance was utilized in the first commercial Li-ion battery, its flammability remained a problem. In rechargeable Li-ion batteries, lithium metal anodes were initially employed; however, safety concerns prompted the creation of an intercalation anode. The lithium graphite electrode was created in 1980 after Rachid Yazami showed that lithium could be electrochemically intercalated in graphite in a reversible manner [3]. However, because liquid solvents co-intercalated with Li+ ions, resulting in the electrode's crumbling and short cycle life, his work was restricted to solid electrolytes (polyethylene oxide). For this, a variety of anode materials were investigated. Akira Yoshino found in 1985 that Li-ions can be reversibly intercalated at a low potential without structural degradation in petroleum coke, a less graphitized carbon [4].

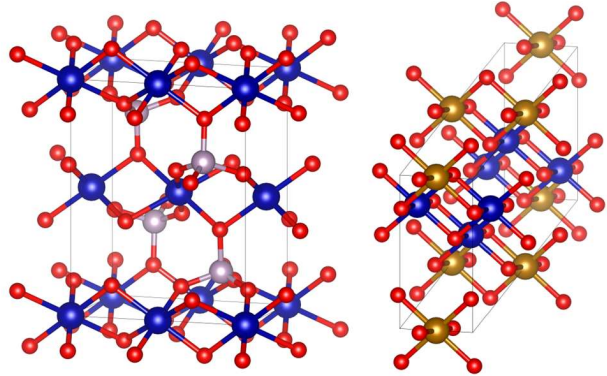

Figure 2. Crystaline Structure of CoPO4 (mp-776270) and Fe(CoO3)2 (mp-761530)

Its amorphous carbon portions function as covalent joints, which gives it stability. Petroleum coke was the first commercial intercalation anode for Li-ion batteries because of its cycle stability, even though it had a smaller capacity than graphite. China now accounts for 75% of the world's lithium-ion battery production capacity, which has increased dramatically since 2010 and reached 28 GWh in 2016 [5]. It is projected that production will range from 200 to 600 GWh in 2021 and 400 to 1,100 GWh in 2023 [6]. For their efforts, John B. Goodenough, Rachid Yazami, and Akira Yoshino were honored with multiple honors, including the 2019 Nobel Prize in Chemistry and the 2012 IEEE Medal for Environmental and Safety Technologies. Additionally, Jeff Dahn has been honored with the International Battery Materials Association's Yeager award and the ECS Battery Division Technology Award.

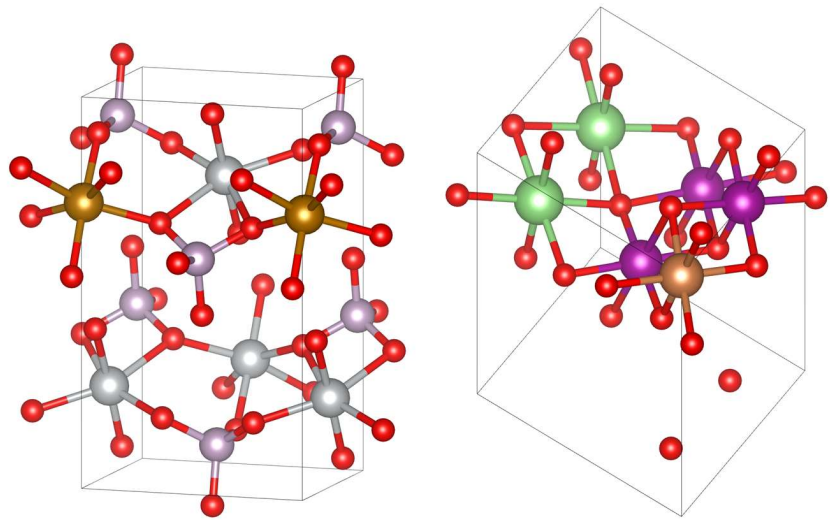

Figure 3. Crystaline Structure of FeNi3(PO4)4 (mp-755553) and Li2Mn3SbO8 (mp-752791)

Depending on voltage and temperature, batteries have a higher open-circuit voltage than aqueous batteries, and internal resistance rises with age and cycling [7]. This lowers the maximum current flow by causing the terminal voltage to drop under load. The battery can no longer sustain typical discharge currents as resistance rises without experiencing an intolerable voltage loss or overheating. Positive and negative electrode lithium-ion batteries typically have a charging voltage of 3.6 V and a nominal open-circuit voltage of 3.2 V. When charging, their maximum voltage is

4.2 V. With current-limiting circuitry, the charging process never stops until 4.2 V is attained. At 3% of the original current, the charge is cut off. Lithium-ion batteries used to take at least two hours to fully charge and were not fast-chargeable.

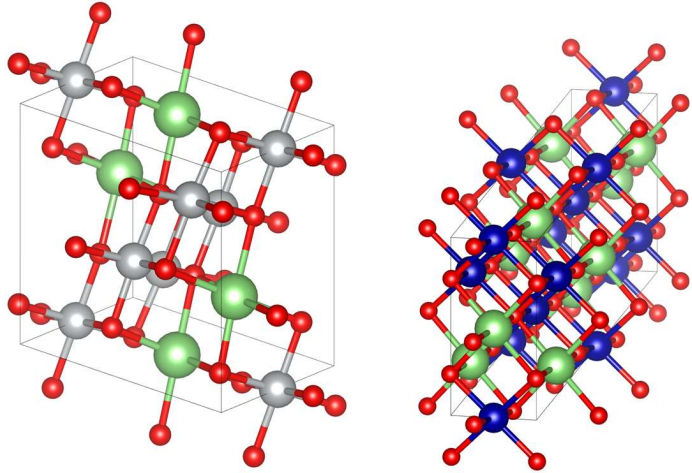

Figure 4. Crystaline Structure of Li3(NiO2)5 (mp-762165) and Li3Cr2CoO6 (mp-753687)

The number of complete charge-discharge cycles required to hit a failure threshold establishes a lithium-ion battery's lifespan; this is commonly known as "cycle life." The number of cycles required to attain 80% of the rated battery capacity is divided to determine this lifespan [8]. The formation of the solid electrolyte interface on the anode causes stored batteries in the charged state to lose capacity and increase cell resistance.

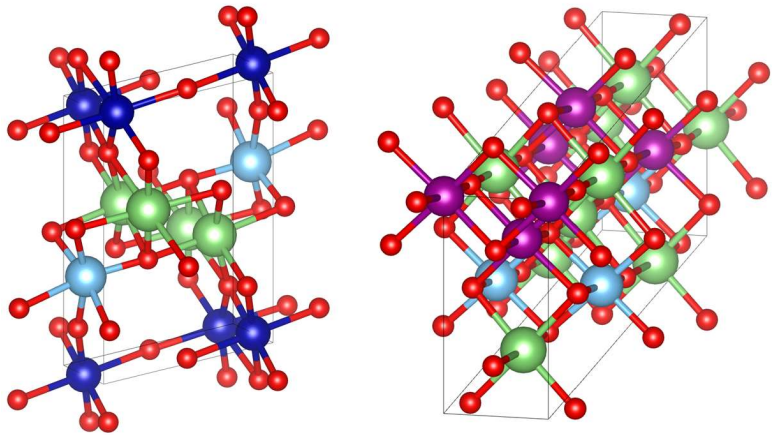

Figure 5. Crystaline Structure of Li3Ti2Co3O10 (mp-753046) and Li3TiMn2O6 (mp-762163)

The full battery life cycle is represented by calendar life, which is affected by a few stressors, including temperature, discharge current, charge current, and state of charge ranges [9]. Full discharge cycles are deceptive since battery life in devices like computers, cellphones, and electric cars are not fully charged and depleted. Cumulative discharge, or the total charge given by the battery over its whole life or equivalent full cycles, is what researchers employ. Temperature and state of charge have an impact on battery degradation during storage; full charge (100% SOC) and high temperature can cause a significant loss in capacity and the production of gas [10].

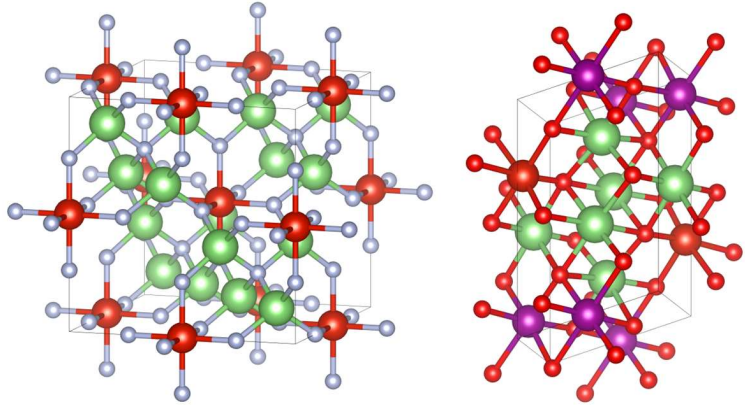

Figure 6. Crystaline Structure of Li4VF7 (mp-753113) and Li5Mn3V2O10 (mp-756863)

The total energy delivered over the battery's life, including charging expenses, is calculated by multiplying the cumulative discharge of the battery by the rated nominal voltage. High temperatures (50–60 °C) are frequently used in lithium-ion battery aging studies to expedite the completion of research. In these circumstances, fully charged lithium-iron phosphate and nickel-cobalt-aluminum cells lose about 20% of their cyclable charge in a period of one to two years. Manganese-based cathodes exhibit quicker degradation (20–50%), whereas anode aging is the most significant degradation route. Lithium-ion battery degradation proceeds along the same pattern at 25 °C, but it happens half as quickly [11].

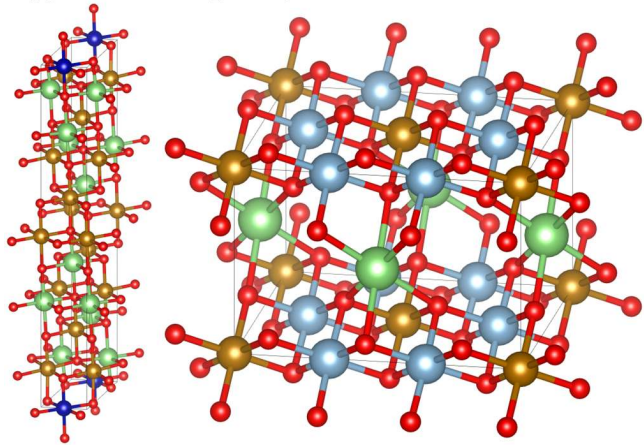

Figure 7. Crystaline Structure of Li8Fe9CoO20 (mp-759442) and LiAl2FeO6 (mp-755069)

According to limited experimental data, lithium-ion batteries lose roughly 20% of their cyclable charge after 1000–2000 cycles at 25°C or three to five years. Titanate anodes have a longer lifespan than graphite anodes because they are not affected by SEI development. But beyond 1000–2000 days, other degradation processes including $Mn^{3+}$ dissolution and $Ni^{3+}/Li^+$ place exchange take place, therefore using titanate anodes doesn't make fuel cells more durable [12]. To fight climate change and lower carbon dioxide emissions, energy-saving and emission-reduction technologies must be developed. The global energy system has been profoundly affected by the energy crisis. Governments, corporations, and organizations pledged to speed up the switch to completely zero-emission vehicles at the COP26 meeting in Glasgow to meet the targets of the Paris Agreement by 2040 or 2035 in major markets [13].

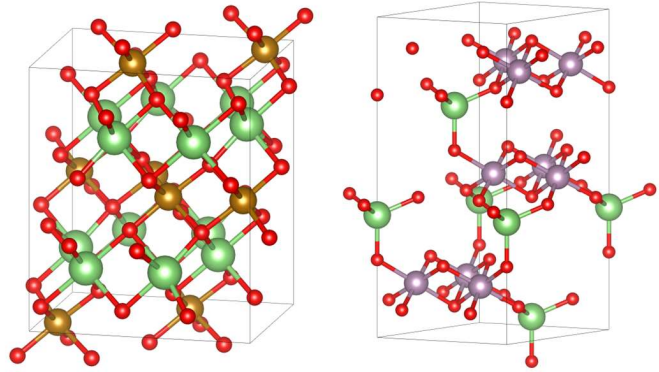

Figure 8. Crystaline Structure of LiFeO3 (mp-752521) and LiMo3O8 (mp-25273)

Since electric vehicles (EVs) run on lithium-ion batteries rather than gasoline or diesel and global EV sales are expected to reach 6.75 million units in 2021, vehicle electrification is a potential strategy to combat the energy crisis and lower carbon dioxide emissions. Despite LiBs' high power, energy density, and lengthy life cycle, the number of electric cars (EVs) sold rose by 108% in 2022. To lower overdesign costs and improve vehicle performance and efficiency, it is essential to maintain a reliable, effective, and safe battery storage system [14].

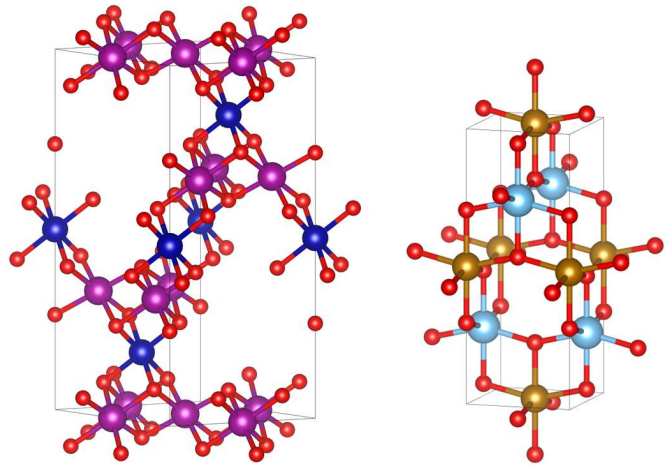

Figure 9. Crystaline Structure of Mn3CoO8 (mp-756482) and TiFeO4 (mp-754900)

Like fuel gauges, electric vehicles (EVs) employ the state of charge as a measure of battery capacity. But it's difficult to estimate because it's not readily measurable. For estimating the state of charge, researchers have created safe and trustworthy techniques such as look-up table methods, ampere-hour integral methods, filter-based methods, observer-based methods, and data-driven approaches. The goal of these techniques is to give a trustworthy and accurate picture of the battery's remaining capacity while a car is in operation. Although the look-up table and ampere-hour integral approaches are simple, they have drawbacks, such as poor robustness in sensor faults and low precision in real-world applications. In contrast, filter-based and observer-based approaches provide improved noise robustness, self-correction, and estimation accuracy [15]. Building a machine learning state of charge prediction method requires gathering data from actual EV operations or battery tests. Before machine learning models are trained, raw data is processed. The accuracy and complexity of the model are taken into consideration while choosing a machine learning algorithm. The chosen algorithm is utilized to estimate state of charge using fresh data following training and validation. For state of charge estimation, many machine learning algorithms are described, such as gaussian process regression (GPR) techniques, support vector machines (SVM), deep learning (DL), and shallow neural networks (NN). Real human brain neurons serve as the inspiration for the potent and alluring machine learning algorithms known as artificial neural networks (ANN). They fall into two categories: deep learning (DL) algorithms and conventional neural networks. One input, hidden, and output layer make up traditional neural networks, sometimes referred to as shallow neural networks. Extreme learning machines (ELM), radio frequency neural networks (RBFNN), backpropagation neural networks (BPNN), and wavelet neural networks (WNN) are a few examples.

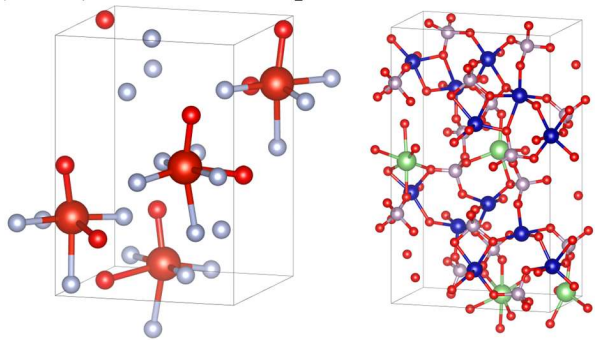

Figure 10. Crystaline Structure of VOF3 (mp-761177) and LiCo3P4O15 (mp-775872)

The goal of materials discovery is to develop effective materials for commercialization, a process that may take up to 20 years. The goal of accelerated techniques is to reach commercialization more quickly. The case of vaccine development, which took less than a year to create and release following the COVID-19 pandemic, can teach the materials science area a lot. This was brought about by a change in technology, such as a 2008 discovery that reduced the cost of sequencing DNA, and an unprecedented level of worldwide research intensity, which enabled researchers to screen more vaccines than previously [16]. Although it has different requirements, machine learning for energy technologies is analogous to machine learning for other domains, such as biomedicine.

2. **Procedure**

Over the past few decades, research has been done on a range of materials and technologies that assist create electronics that can be useful in applications for electric vehicles [17–29]. Here Quantum Espresso software has been used to calculate the magnetic properties of several potential Li-ion battery materials (Table 1). Crystal structures of these materials are shown in Figure 1 to Figure 10.

| Material | Total Energy | Total Magnetization |
| --- | --- | --- |
| LiCuO (mp-5127) | -2128.59389754 Ry | Non-magnetic |
| Al3Cr3-SbO8 (mp-756662) | -3341.37529193 Ry | 10.00 Bohr mag/cell |
| CoPO4 (mp-776270) | -1908.36808563 Ry | 14.82 Bohr mag/cell |
| Fe(CoO3)2 (mp-761530) | -2340.40702394 Ry | 8.15 Bohr mag/cell |
| FeNi3(PO4)4 (mp-755553) | -2071.37795011 Ry | 8.45 Bohr mag/cell |
| Li2Mn3SbO8 (mp-752791) | -1176.39885635 Ry | 11.62 Bohr mag/cell |
| Li3(NiO2)5 (mp-762165) | -2170.23688619 Ry | 2.79 Bohr mag/cell |
| Li3Cr2CoO6 (mp-753687) | -1879.64734287 Ry | 6.14 Bohr mag/cell |
| Li3Ti2Co3O10 (mp-753046) | -1588.85354319 Ry | 6.56 Bohr mag/cell |
| Li3TiMn2O6 (mp-762163) | -1664.77928888 Ry | 17.97 Bohr mag/cell |
| Li4VF7 (mp-753113) | -2174.02743785 Ry | 8.00 Bohr mag/cell |
| Li5Mn3V2O10 (mp-756863) | -1408.15051582 Ry | 15.95 Bohr mag/cell |
| Li8Fe9CoO20 (mp-759442) | -4179.79937119 Ry | 19.62 Bohr mag/cell |
| LiAl2FeO6 (mp-755069) | -1332.05129369 Ry | 6.18 Bohr mag/cell |
| LiFeO3 (mp-752521) | -3718.79128032 Ry | 23.28 Bohr mag/cell |
| LiMo3O8 (mp-25273) | -2266.37815043 Ry | 7.66 Bohr mag/cell |
| Mn3CoO8 (mp-756482) | -3782.73857024 Ry | 30.75 Bohr mag/cell |
| TiFeO4 (mp-754900) | -1222.08046773 Ry | 7.26 Bohr mag/cell |
| VOF3 (mp-761177) | -1328.04415756 Ry | -0.99 Bohr mag/cell |
| LiCo3P4O15 (mp-775872) | -6328.13428109 Ry | 8.72 Bohr mag/cell |

Table 1. Total energy and total magnetization Li-ion battery materials

### 3. Conclusion

In conclusion, this research paper provides magnetic properties of several potential Li-ion battery materials and a comprehensive overview of the advancements in Li-ion battery technology, addressing safety concerns, operational characteristics, lifespan considerations, and the role of Li-ion batteries in combating climate change.

**Acknowledgements**

This work was supported as part of the Modeling and Simulation Program (MSP) grant funded by the US Department of Education under Award No. P116S210002 and Improving Access to Cyber Security Education for Underrepresented Minorities funded by the US Department of Education under Award No. P116Z230007.